# Agents for Agents: An Interrogator-Based Secure Framework for Autonomous Internet of Underwater Things

Ali Akarma [1,*], Toqeer Ali Syed [1], Abdul Khadar Jilani [2], Salman Jan [3], Hammad Muneer [4], Muazzam A. Khan [5] and Changli Yu [6]

1. AI Center, Faculty of Computer and Information Systems, Islamic University of Madinah, Madinah 42351, Saudi Arabia; 443059463@stu.iu.edu.sa (A.A.); toqeer@iu.edu.sa (T.A.S.)
2. College of Computer Studies, University of Technology Bahrain, Sitra 18041, Bahrain; a.jilani@utb.edu.bh (A.K.J.)
3. Faculty of Computer Studies, Arab Open University-Bahrain, Manama 32038, Bahrain. salman.jan@aou.org.bh (S.J.)
4. Department of Computer Science, The Islamia University of Bahawalpur, Bahawalpur 63100, Pakistan. S21bdocs1m01009@iub.edu.pk (H.M.)
5. ICESCO Chair in Big Data Analytics and Edge Computing, Quaid-e-Azam University, Islamabad 45320, Pakistan; muazzam.khattak@qau.edu.pk (M.A.K.)
6. Harbin Institute of Technology, Weihai 264209, China; yuchangli@hitwh.edu.cn (C.Y.)

* Correspondence: 443059463@stu.iu.edu.sa

**Abstract**

Autonomous underwater vehicles (AUVs) and sensor nodes increasingly support decentralized sensing and coordination in the Internet of Underwater Things (IoUT), yet most deployments rely on static trust once authentication is established, leaving long-duration missions vulnerable to compromised or behaviorally deviating agents. In this paper, an interrogator-based structure is presented that incorporates the idea of behavioral trust monitoring into underwater multi-agent operation without interfering with autonomy. Privileged interrogator module is a passive communication metadata analyzer that uses a lightweight transformer model to calculate dynamic trust scores, which are used to authorize the forwarding of mission-critical data. Suspicious agents cause proportional monitoring and conditional restrictions, which allow fast containment and maintain network continuity. The evidence of trust is stored in a permissioned blockchain consortium which offers identity management which is not tampered and is decentralized without causing the overhead of public consensus mechanisms. Simulation-based analysis shows that the evaluation of the result compares to a relative improvement of 21.7% in the detection accuracy compared to the static trust baselines with limited energy overhead. These findings suggest that behavior driven validation has the capability of reinforcing underwater coordination without compromising scalability and deployment.

## 1. Introduction

Internet of Underwater Things (IoUT) has become a paradigm of large-scale sensing, exploration, and monitoring of aquatic environments. IoUT systems facilitate environmental monitoring, maritime security, infrastructure inspection, and disaster prediction through the underwater sensor nodes, autonomous underwater vehicles (AUVs), and surface gateways by connecting them through the internet [1]. Nevertheless, in contrast to deployments on land, underwater networks depend on acoustic communication channels, which are defined by low bandwidth, large latency, intermittent connection, and a strict energy limit [2]. Such conditions have a strong effect on the design of coordination and security systems in underwater multi-agent systems.

The recent development of decentralized multi-agent intelligence allows the underwater agents to coordinate sensing, navigation, and relaying of data with minimum human intervention [3]. The strategies of learning increase adaptiveness and scalability in the case of partial observability but most of the current models implicitly expect that agents can be trusted once they are authenticated during deployment. This fixed trust model presents serious vulnerabilities in the long-term missions, whereby agents can be compromised, spoofed, malfunctioning or have behavioral inconsistencies under the influence of adversary, or internal flaws [4].

Lightweight cryptographic authentication and key management are the main focus of the traditional underwater security mechanisms. Although these methods are effective in preventing outside intrusion, they offer weak insight into the insider threat or gradual change in behavior of agents who still have valid credentials [5]. Ongoing intensive cryptography surveillance is also limited by power and communication constraints [6], which drives the urge to develop dynamic, behavior-based security that maintains decentralized autonomy.

To overcome this issue, we offer an interrogator-based model that implements the idea of constant behavioral trust checking into the operational process of underwater agents. A specialized interrogator module is able to monitor non-invasively, by passively monitoring the metadata of communications and temporal patterns of interaction, but not by analyzing the contents of payloads. A lightweight transformer model is used to compute dynamic trust scores that are used to control authorization of mission-critical data forwarding.

Identity records and trust evidence are stored in a consortium of permissioned blockchain comprised of surface gateways and validator nodes, which constitute a tamper-resistant trust ledger that implements zero-trust participation [7]. In the case of having anomalous behavior, the framework goes into a proportional interrogation state where the intensity of monitoring goes up and remedial actions, including communication throttling or temporary isolation are implemented via decentralized consensus [8].

In contrast to previous IoUT trust architectures, the suggested framework directly co-designs behavioral inference with acoustic communication constraints and thus is able to keep continually validating the workflow without interrupting bandwidth-constrained coordination processes. The proposed solution is more resilient against insider compromise than conventional authentication, as it replaces the concept of static authentication with the idea of continuous behavioral validation, without compromising operational autonomy and scalability. Figure 1 represents the architecture of system with the emphasis on lack of connectivity between the operational data flow, behavioral trust enforcement and decentralized governance.

The inability to continuously verify agent action can be disastrous to the operations in underwater missions. For example, the compromised relay node can affect the routing path quietly, leading to important sensory information being dropped or bumped at the time when carrying out time-intensive processes like infrastructure inspection or environmental danger surveillance. Similarly, the spoofed coordination messages can propagate quickly with the decentralized networks and impair the mission reliability before anomalies are detected using cryptographic mechanisms. These dangers emphasize that solely static authentication is not adequate in the case of long-term autonomous deployments, which encourages ongoing behavioral verification. The primary contributions of this work are:

- An interrogator-based architecture enabling continuous behavioral trust verification in decentralized IoUT deployments.
- A lightweight transformer-driven inference mechanism designed for sparse acoustic communication patterns.
- A permissioned governance model with tiered enforcement that balances rapid local response with decentralized validation.

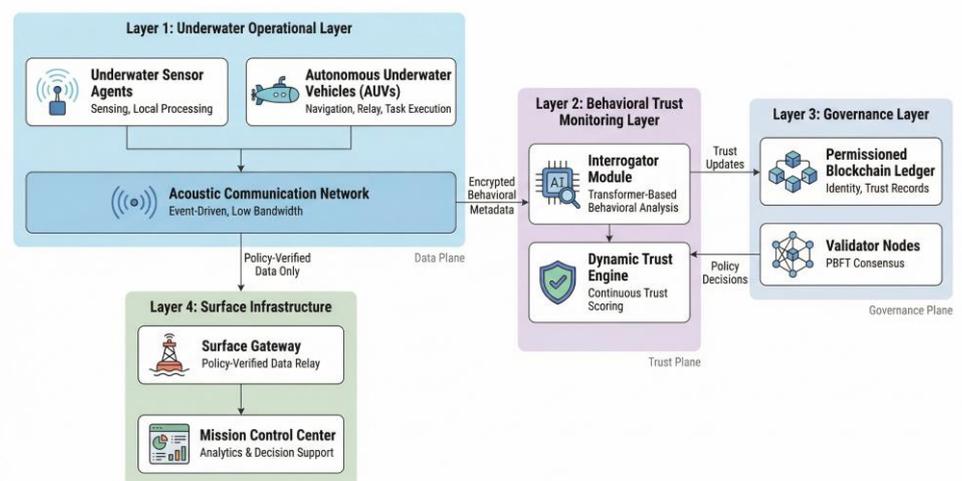

**Figure 1.** System architecture of the proposed agent-for-agents framework for the Internet of Underwater Things (IoUT), illustrating operational underwater agents, the interrogator-based security monitoring layer, permissioned blockchain consortium governance, surface control infrastructure, and external administrative stakeholders.

## 2. Related Work

The management of security and trust have become important in distributed Internet of Things (IoT) and multi-agent settings since autonomous systems are being operated with less and less supervision. Current literature covers behavioral trust modeling, underwater network security, blockchain-based governance, and temporal anomaly detection but these efforts have mostly been developed independently, creating gaps in the area of long-term autonomous underwater deployments.

Trust mechanisms that are based on behavior have been extensively studied to deal with insider threats which overcome conventional authentication. Zhang et al. [9] presented a dynamic trust model that relies on the history of interactions, whereas Bao and Chen [10] incorporated social trust as well as quality-of-service in enhancing resilience in wireless sensor networks. Machine learning has also boosted trust inference, with Liu et al. [11] used the anomaly detection technique to detect behavioral deviations and Guo et al. [12] used reinforcement learning to perform adaptive trust calibration. Nonetheless, these methods majorly assume a fixed terrestrial connectivity and do not take into consideration the harsh bandwidth and latency limitations of acoustic conditions.

Underwater wireless networks have their own set of challenges such as long propagation delays, low throughput and high cost of energy. Zhou et al. [13] presented lightweight cryptography mechanisms to be used in underwater sensors, while Akyildiz et al.

[14] emphasized on the necessity of adaptive defenses in dynamic acoustic topologies. However, the current frameworks mostly focus on authentication and key management as opposed to continuous verification of behavior, which is implicitly based on the assumption that authenticated nodes are still trustworthy during the mission lifecycles.

Blockchain has become a promising instrument of tamper resistant coordination of distributed systems. Khan et al. [15] have shown that permissioned ledgers are suitable to use in resource-constrained IoT deployments, and Reyna et al. [16] have demonstrated that consortium architectures could offer auditable trust management with limited latency. However, when it comes to underwater environments, it is still difficult to implement devolved governance because of ad hoc connectivity and acoustic latencies. Temporal models such as Long Short-Term Memory networks [17] has performed well in detecting behavioral anomalies through sequence modeling, with recent work emphasizing lightweight inference for constrained platforms.

Regardless of these developments, there is limited literature that combines behavioral trust, security constraints underwater, decentralized governance and temporal inference in one architecture [18]. The current solutions are based on either fixed authentication or do not provide scalable mechanisms to perform a continuous assessment of trust. To fill this gap, this paper presents an interrogator-based framework which decouples behavioral verification and operational autonomy and uses lightweight transformer inference and permissioned blockchain governance to permit non-intrusive monitoring and decentralized enforcement in the deployment of IoUTs [19-21]. In contrast to earlier IoUT security solutions which depend mainly on cryptographic authentication, or fixed-point reputation models [22], the suggested framework incorporates the ongoing behavioral verification in the core of decentralized coordination processes and explicitly considers acoustic communication limitations.

## 3. Methodology

This section presents the interrogator-based security architecture for autonomous Internet of Underwater Things (IoUT) deployments. As illustrated in Fig. 1, the framework combines operational actors, behavioral trust tracking layer and a permissioned governance infrastructure as part of an observe- decide-act paradigm. The design maintains decentralized autonomy with the capability to verify the trust continuously without interruption and without violating acoustic delivery and energy limitations.

*3.1. System Model*

Let A = $\{a_1, a_2, \ldots, a_N\}$ denote a heterogeneous set of underwater agents consisting of sensor nodes and autonomous underwater vehicles (AUVs) operate in an environment characterized by high latency, limited bandwidth, and intermittent connectivity. At time step $t$, agent $a_i$ observes state $s_i^t$ capturing environmental measurements, neighbor context, link quality, and residual energy, and selects an action $u_i^t$ according to decentralized policy $\pi_i(u_i^t|s_i^t)$.

Agents aim to maximize cumulative expected reward over mission horizon $T$:

$$\max_{\pi_i} E\left[\sum_{t=0}^{T} \gamma^t r_i(s_i^t, u_i^t)\right],$$

where $r_i(\cdot)$ encodes sensing fidelity, routing reliability, and energy sustainability.

*3.2. Deployment Assumptions and Operational Pipeline*

The architecture presupposes the implementation of the heterogeneous compute capabilities: resource-constrained sensors are used to carry out lightweight communication, whereas more capable platforms (e.g., Jetsonclass devices on AUVs or gateways) are used to do behavioral inference and ledger participation. The interrogator functionality is a privileged software module deployed on the selected nodes without the need of deploying extra hardware. Operational agents have an observe-decide-act cycle where sensory, decentralized thinking, and event-based acoustic messages are used to assist in navigation and relay work in a cooperative manner. Energy consumption is modeled as

$$E_{tx}(l,d) = lE_{elec} + l\varepsilon_{amp}d^k, \quad E_i^t = E_i^{sense} + E_i^{compute} + E_{tx}(l,d),$$

enabling adaptive duty cycling for long-duration missions.

*3.3. Behavioral Trust Inference*

One of the interrogator modules is passive in that it monitors metadata of communication but does not examine the payload content. A behavioral sequence of each agent captures the features like inter packet timing, routing stability, retransmission rates, neighbor churn and protocol compliance.

$$X_i = \{x_i^1, \dots, x_i^K\}$$

The sequences are fed through a small transformer encoder with four attention layers and hidden dimension 128 (with 1.2M parameters after quantization). The model is able to capture the long-range temporal dependencies in sparse acoustic traffic as well as efficient edge inference. Trust is estimated as

$$\tau_i^t = f_\theta(X_i^t), \quad \tau_i^t \in [0,1],$$

and updated via exponential smoothing:

$$\tau_i^{t+1} = \alpha\tau_i^t + (1-\alpha)f_\theta(X_i^t).$$

Although the lightweight payload sketching is still possible, metadata analysis is given priority to maintain the energy efficiency and the communication overhead in bandwidth limited acoustic environment is reduced to a minimum.

3.3.1. Computational Feasibility

The quantized transformer can be run on the computational envelope of embedded GPU platforms at fixed monitoring rates as opposed to running the algorithm in continuous mode. Interrogator modules are placed in nodes with higher capacity, so that sensing agents are not affected by the inference overhead. A quantized LSTM baseline was more stable in the case of sparse acoustic traffic, which encouraged the choice of transformer. A quantized LSTM baseline exhibited reduced stability under sparse acoustic traffic, motivating the transformer selection. The quantized model requires approximately $6-8$ MB of memory and executes inference in about $120-180$ ms on Jetson-class embedded GPUs operating within a $5-10$ W power envelope, remaining compatible with the energy budgets of higher-capability AUV platforms.

*3.4. Governance and Enforcement*

Trust evidence is stored in a permissioned ledger that is run in surface gateways and a few validators through Practical Byzantine Fault Tolerance (PBFT). Due to the nature of consortium consensus which controls continuous policy changes instead of safety-critical control loops, the corresponding latency does not cause interference with time-sensitive underwater maneuvers like collision avoidance or emergency surfacing. Trust updates are only committed in summary mode and this limits the communication overhead. Agents are authorized to forward mission-critical data only when $t_i \geq T_{min}$ otherwise triggering an interrogation state.

3.4.1. Computational Feasibility

In order to reduce the consensus latency, the framework uses tiered enforcement. On the detection of high confidence anomalies, interrogators implement immediate local constraints, e.g. routing exclusion or transmission throttling whereas policy updates are controlled by consortium validation. The hybrid model limits the enforcement latency to the period of monitoring and maintains the guarantee of decentralized trust.

3.4.2. Incident Response Workflow

Upon confirmed behavioral deviation, the interrogator initiates a staged incident response process. The affected agent is first subjected to temporary routing exclusion to prevent propagation of corrupted data while preserving network connectivity. Neighboring agents recompute relay paths using decentralized routing policies, ensuring mission continuity. If anomalous behavior persists across successive monitoring intervals, the consortium enforces logical isolation until recovery procedures or operator intervention occur. This recovery-aware isolation strategy enables rapid containment while minimizing disruption to autonomous coordination.

3.4.3. Operational Modes

The architecture has two operational modes, which are baseline autonomous coordination and security elevated monitoring. In an ideal scenario, when the authenticated agents are involved, the workflow is sensed and routed without interruption as the interrogators run on low duty cycles. Once the trust scores drop to an amount that is less than the predefined thresholds, the system switches to an interrogation state where there is an increased monitoring, cross-validation, and proportional enforcement measures such as routing exclusion or temporary isolation. This design maintains continuity in its operations and allows fast containment of the compromised nodes.

*3.5. Threat Model*

Attackers can also steal authentic agents, inject counterfeit data, drop packets, replay their transmissions, or organize possible insider attackers and maintain valid credentials. Attack on the nodes of the interrogator is addressed by using redundant monitoring and cross-verification of trust updates. Even though adversarial manipulation of behavioral features has not been solved yet, temporal interaction patterns make the process of stealthy evasion even more challenging without interfering with the operational consistency. The metadata mimicry, however, cannot be synchronized unless the protocol level consistency is maintained across several windows in time which raises the cost of adversarial operation and decreases the stealth success.

*3.6. Adaptive Security Enforcement*

The level of security actions is dependent on the perceived risk. Interrogator is used in the low duty cycles when the conditions are stable and increases the surveillance when anomalies arise. The architecture allows maintaining the scalability by separating operational control and trust evaluation and provides the ability to conduct continuous verification at limited energy cost. Additional reduction on the false positives caused by temporary effects on the environment like mobility changes or channel variations is done through exponential smoothing and persistence checks.

## 4. Prototype Implementation

The NS-3 Aqua-Sim simulator was used to test a prototype of the interrogator-based framework using bandwidth-limited and high-latency acoustic conditions. This deployment involved 50 heterogeneous agents distributed in a 1 $km^2$ area with both the presence of stationary sensors and mobile AUV relays adhering to a limited random waypoint mobility model. Acoustic communication was at 10-20 kbps and channel propagation delays were around 0.67 s/km, and event-based messaging minimized channel contention and had an appearance of duty-cycled underwater operation.

Lightweight decentralized coordination policies were implemented by operational agents that were trained offline and inferred at low cost. The standard underwater transmission models were applied in energy consumption, where the remaining battery levels were utilized in adaptive duty cycling.

Interrogator functionality was implemented on high capacity nodes that modeled edge enabled AUVs or surface gateways. A set of modules monitored metadata of communication, such as inter-packet timing variance, retransmission frequency, routing stability, neighbor churn and protocol deviations and analyzed behavioral sequences by the quantized transformer in Section 3. The computation of trust updates was performed at 30 s intervals, which allowed finding anomalies in the near-real-time without looking at the channels continuously. The quantized transformer requires approximately 0.35-0.6 J per inference on embedded GPU-class devices, corresponding to less than 7% of the estimated daily energy budget of higher-capability AUV platforms.

Trust summaries were saved into an authorized network based on Hyperledger Fabric that had nine validators running PBFT consensus. Surface infrastructure had a role of a validator to reduce the overhead of underwater communication and aggregated trust deltas and confirmed security events are only logged to limit transaction volume. The selective packet drops, manipulating routes, abnormal transmission bursts and coordinated insider behavior among the compromised nodes were introduced as adversarial behaviors in 15% of the agents. To support experimental reproducibility and clarify deployment assumptions, the primary simulation parameters are summarized in Table 1.

**Table 1.** Simulation parameters for the interrogator-based IoUT framework.

| | | |
|---|---|---|
| **Network Parameters** | Simulator | NS-3 Aqua-Sim |
| | Network Size | 50 agents |
| | Deployment Area | 1 km$^2$ |
| | Acoustic Rate | 10–20 kbps |
| **Security Configuration** | Monitoring Interval | 30 s |
| | Adversarial Nodes | 15% |
| | Validators | 9 (PBFT) |
| **Model Parameters** | Sequence Length | 64 |
| | Transformer Parameters | 1.2M |
| | Trust Threshold ($\tau_{min}$) | 0.65 |
| | Smoothing Factor ($\alpha$) | 0.8 |
| | Interrogator Inference Energy | 0.35-0.6 J per inference |
| | Estimated Daily Overhead | <7% of AUV energy budget |

## 5. Performance Evaluation

The proposed framework was compared with two baselines: (i) a static-trust architecture whereby authenticated agents are allowed to act without behavioral validation and (ii) a Bayesian reputation model which is probabilistic trust inference. Measures are the performance of anomaly detection, energy overhead and communication reliability. All the measurements are averaged over 30 simulation runs.

*5.1. Trust Inference Performance*

The accuracy of anomaly detection with respect to monitoring intervals is illustrated inv Fig. 2. The interrogator-based framework recorded a mean accuracy of 94.2%, which was higher than the case with static trust 72.5% and Bayesian trust 86.1%. The precision and the recall were 0.92 and 0.95, respectively, which showed robust discrimination between legitimate and compromised agents. The latency of early-stage detection was also reduced by about 27% and the time of malicious nodes to affect routing behavior was minimized.

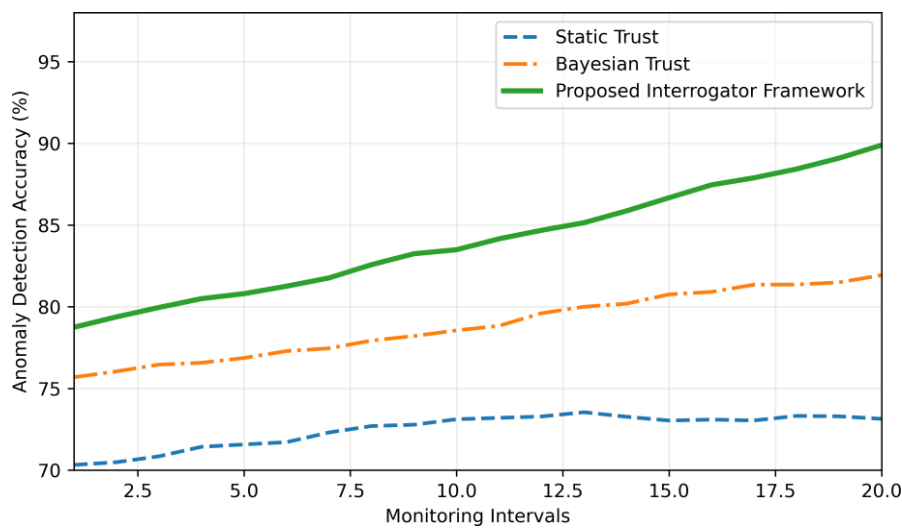

**Figure 2.** Anomaly detection accuracy comparison across trust architectures. The proposed interrogator-based framework achieves higher detection performance than static and Bayesian trust models while maintaining stable convergence across monitoring intervals.

*5.2. Energy Overhead*

The trends in residual energy in Fig. 3 show that behavioral monitoring will cause only an average overhead of 5.8% as compared to the case of a static trust, which is within manageable limits in the case of long-duration missions. The ability to control the activation of the interrogator modules has indicated that abrupt depletion can be avoided and as such the continuous verification of trusts can be brought about without affecting the operations life.

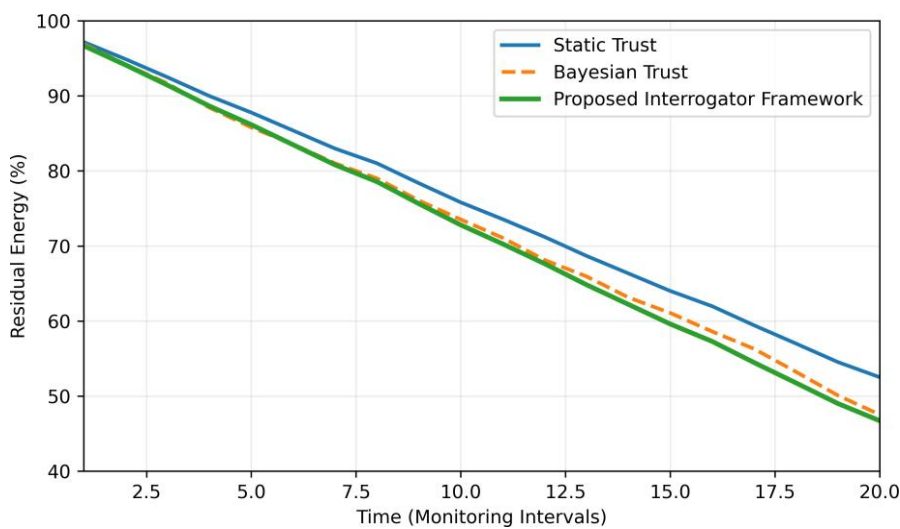

**Figure 3.** Residual energy comparison across trust architectures. The proposed interrogator-based framework introduces a modest average overhead of 5.8% relative to static trust while maintaining energy levels suitable for long-duration underwater missions.

*5.3. Communication Reliability*

The ratio of packet delivery increased to 91.6%, which is higher than 79.4% in the case of the static trust and 86.7% in case of the Bayesian baseline, as depicted in Fig. 4. Such improvements lead to the fact that proactive detection and separation of suspicious agents stabilize routing paths and minimum retransmissions. Notably, the extra overhead of monitoring does not affect the stability of the performance of the delivery, which means that routing dynamics are not interfered with by the enforcement of trust.

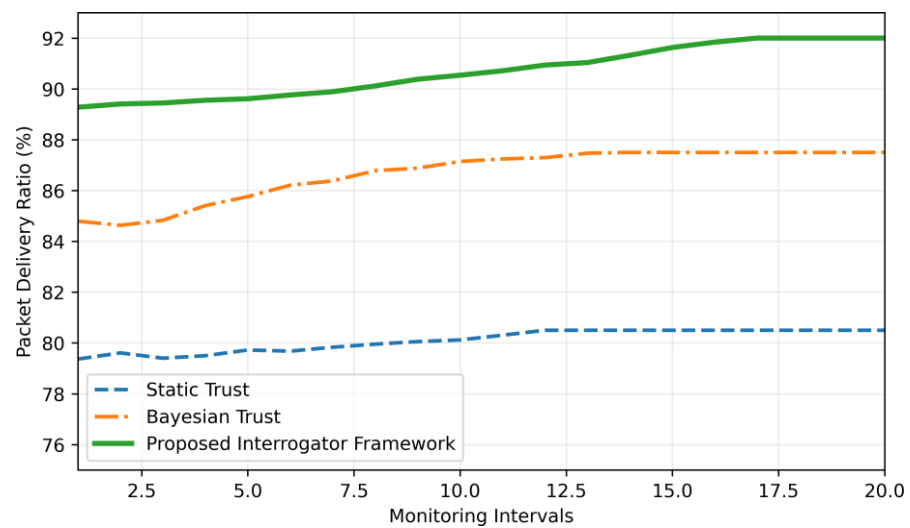

**Figure 4.** Packet delivery ratio comparison across trust architectures under adversarial conditions. The proposed framework maintains higher routing stability by proactively isolating anomalous agents.

*5.4. Scalability*

Accuracy of the detection was more than 92% at 50 agents with no apparent growth in the enforcement latency indicating that the monitoring architecture grows with a moderate growth in deployment density. Summarized trust commits continued to limit the ledger traffic. As much as simulation based validation is a useful tool that allows controlled experimentation, deployment to real physical underwater platforms will be explored in the future.

## 6. Conclusion

This paper proposed an Internet of Underwater Things (IoUT) interrogator-based security architecture that combines multi-agent autonomy, which is decentralized, with constant behavioral trust validation. The framework improves the resilience of operation in the face of compromised agents by integrating dynamic trust assessment at the operational lifecycle, whereas maintaining scalable coordination. Simulation based testing and prototype implementation showed that it had better accuracy in detecting anomalies, limited energy overhead and greater communication reliability compared to the static and probabilistic trust baselines. These findings suggest that, based on the behavior-based validation, it is possible to significantly decrease the operational impact of malicious nodes without interrupting decentralized workflows. The architecture also includes a governed layer which allows tamper resistant trust management and does not involve the computational overhead of public blockchain mechanisms. Monitoring and enforcement are proportionally activated to ensure a long-term operation in underwater circumstances of bandwidth constraints and energy limitation. Some limitations that are related to this study are simulation-based validation and moderate deployment scale. Future research will examine the practical implementation of a platform integrating lightweight algorithms, the optimization of lightweight sequence models of ultra-low-power systems, and adaptation to concurrence of highly intermittent acoustic links to consensus. Comprehensively, the offered solution suggests that the behavioral verification that is performed continuously will be able to substitute the assumptions of the generally static trusts without compromising the efficiency of the operations, thus establishing a viable base of the secure and robust network of the IoUT deployments in the adversarial maritime conditions.

## References

1. Mohsan, S.A.H.; Li, Y.; Sadiq, M.; Liang, J.; Khan, M.A. Recent advances, future trends, applications and challenges of internet of underwater things (IoUT): A comprehensive review. J. Mar. Sci. Eng. 2023, 11, 124. https://doi.org/10.3390/jmse11010124
2. Nkenyereye, L.; Nkenyereye, L.; Ndibanje, B. Internet of underwater things: A survey on simulation tools and 5G-based underwater networks. Electronics 2024, 13, 474. https://doi.org/10.3390/electronics13030474
3. Jan, S.; Syed, T.A.; Ali, G.; Akarma, A.; Belgaum, M.R.; Ali, A. Agentic AI framework for individuals with disabilities and neurodivergence: A multi-agent system for healthy eating, daily routines, and inclusive well-being. arXiv 2025, arXiv:2511.22737. https://doi.org/10.48550/arXiv.2511.22737
4. Hua, M.; Qi, X.; Chen, D.; Jiang, K.; Liu, Z.E.; Sun, H.; Zhou, Q.; Xu, H. Multi-agent reinforcement learning for connected and automated vehicles control: Recent advancements and future prospects. arXiv 2025, arXiv:2312.11084. https://doi.org/10.48550/arXiv.2312.11084
5. Sliwa, J.; Wrona, K.; Shabanska, T.; Solmaz, A. Lightweight quantum-safe cryptography in underwater scenarios. In Proceedings of the 2023 IEEE 48th Conference on Local Computer Networks (LCN), Sydney, Australia, 13–16 November 2023; IEEE: Piscataway, NJ, USA, 2023; pp. 1–6. https://doi.org/10.1109/LCN58197.2023.10223321
6. Gupta, M.; Gera, P.; Mishra, B. A lightweight certificateless signcryption scheme based on HCC for securing underwater wireless sensor networks (UWSNs). In Proceedings of the 2023 16th International Conference on Security of Information and Networks (SIN), Hyderabad, India, 18–20 December 2023; IEEE: Piscataway, NJ, USA, 2023; pp. 1–8. https://doi.org/10.1109/SIN60469.2023.10474770
7. Jan, S.; Razzaqi, H.A.; Akarma, A.; Belgaum, M.R. A blockchain-monitored agentic AI architecture for trusted perception-reasoning-action pipelines. arXiv 2025, arXiv:2512.20985. https://doi.org/10.48550/arXiv.2512.20985
8. Shi, W.; Ling, Q.; Wu, G.; Yin, W. EXTRA: An exact first-order algorithm for decentralized consensus optimization. arXiv 2014, arXiv:1404.6264. https://doi.org/10.48550/arXiv.1404.6264.
9. Zhang, P.; Jiang, C.; Pang, X.; Li, Y. A dynamic trust evaluation model of user behavior based on transformer. *Int. J. Netw. Secur.* **2022**, *24*, 984–993. https://doi.org/10.6633/IJNS.20221124(6).02
10. Bao, F.; Chen, R.; Chang, M.; Cho, J.H. Hierarchical trust management for wireless sensor networks and its applications to trust-based routing and intrusion detection. *IEEE Trans. Commun.* **2012**, *60*, 169–183. https://doi.org/10.1109/TCOMM.2012.031912.110179
11. Liu, Y.; Garg, S.; Nie, J.; Zhang, Y.; Xiong, Z.; Kang, J.; Hossain, M.S. Deep anomaly detection for time-series data in industrial IoT: A communication-efficient on-device federated learning approach. *IEEE Internet Things J.* **2020**, *7*, 3766–3776. https://doi.org/10.1109/JIOT.2020.3011726
12. Guo, W.; Xiong, N.N.; Chao, H.; Hussain, S.; Chen, G. Adaptive trust management based on reinforcement learning for the Internet of Things. *Sensors* **2023**, *23*, 4751. https://doi.org/10.3390/s23104751
13. Zhou, Z.; Xu, C.; Liu, B.; Chen, Y. A lightweight authentication and key agreement scheme for underwater wireless sensor networks. *J. Mar. Sci. Eng.* **2024**, *12*, 831. https://doi.org/10.3390/jmse12050831
14. Akyildiz, I.F.; Pompili, D.; Melodia, T. Underwater acoustic sensor networks: Research challenges. *Ad Hoc Netw.* **2005**, *3*, 257–279. https://doi.org/10.1016/j.adhoc.2005.01.004

15. Khan, A.A.; Laghari, A.A.; Rashid, M.; Li, H.; Gadekallu, T.R. Blockchain at the edge: Performance of resource-constrained IoT networks. *IEEE Trans. Parallel Distrib. Syst.* **2021**, *32*, 7856–7865. https://doi.org/10.1109/TPDS.2020.3013892
16. Reyna, A.; Martín, C.; Chen, J.; Soler, E.; Díaz, M. On blockchain and its integration with IoT: Challenges and opportunities. *Future Gener. Comput. Syst.* **2018**, *88*, 173–190. https://doi.org/10.1016/j.future.2018.05.046
17. Hochreiter, S.; Schmidhuber, J. Long short-term memory. *Neural Comput.* **1997**, *9*, 1735–1780. https://doi.org/10.1162/neco.1997.9.8.1735
18. Syed, T.A.; Alshahrani, A.; Ullah, A.; Akarma, A.; Khan, S.; Nauman, M.; Jan, S. FinAgent: An agentic AI framework integrating personal finance and nutrition planning. *arXiv* **2025**, arXiv:2512.20991. https://doi.org/10.48550/arXiv.2512.20991
19. Alharbi, S. H., Alzahrani, A. M., Syed, T. A., & Alqahtany, S. S. (2024). Integrity and privacy assurance framework for remote healthcare monitoring based on IoT. *Computers*, *13*(7), 164. https://doi.org/10.3390/computers13070164
20. Jan, S., Ali, T., Alzahrani, A., & Musa, S. (2018). Deep convolutional generative adversarial networks for intent-based dynamic behavior capture. *Int. J. Eng. Technol*, *7*(4), 101-103. [Crossref]
21. Ismail, R., Syed, T. A., & Musa, S. (2014, January). Design and implementation of an efficient framework for behaviour attestation using n-call slides. In *Proceedings of the 8th International Conference on Ubiquitous Information Management and Communication* (pp. 1-8). https://doi.org/10.1145/2557977.2558002
22. Syed, T. A., Jan, S., Siddiqui, M. S., Alzahrani, A., Nadeem, A., Ali, A., & Ullah, A. (2022). CAR-tourist: An integrity-preserved collaborative augmented reality framework-tourism as a use-case. *Applied Sciences*, *12*(23), 12022. https://doi.org/10.3390/app122312022